\documentclass[prl,aps,a4paper,showpacs,superscriptaddress,twocolumn]{revtex4}
\usepackage{amsmath,bm}
\usepackage{amsfonts}
\usepackage{amssymb}
\usepackage{graphicx}
\usepackage{dcolumn}
\usepackage{bm}

\begin{document}
\title{Strong thermalization of a mesoscopic two-component Bose-Hubbard model}
\author{J.~M. Zhang}
\affiliation{Beijing National Laboratory for Condensed Matter
Physics, Institute of Physics, Chinese Academy of Sciences, Beijing
100080, China} 
\affiliation{FOCUS Center and MCTP, Department of Physics, University of Michigan, Ann Arbor, Michigan 48109, USA} 
\author{C. Shen}
\affiliation{FOCUS Center and MCTP, Department of Physics, University of Michigan, Ann Arbor, Michigan 48109, USA}
\author{W.~M. Liu}
\affiliation{Beijing National Laboratory for Condensed Matter
Physics, Institute of Physics, Chinese Academy of Sciences, Beijing
100080, China}
%\author{L. M. Duan}
%\affiliation{FOCUS Center and MCTP, Department of Physics, University of Michigan, Ann Arbor, Michigan 48109, USA}

\begin{abstract}
We study thermalization of a two-component Bose-Hubbard model by exact diagonalization. Initially the two components do not interact and are each at equilibrium but with different temperatures. As the on-site inter-component interaction is turned on, perfect thermalization occurs. Remarkably, not merely those simple ``realistic'' physical observables thermalize but even the density matrix of the \textit{whole} system---the time-averaged density matrix of the system can be well approximated by that of a canonical ensemble. A conjecture about this fact is put forward.
\end{abstract}
 \maketitle

For an isolated system, how and in which sense thermal equilibrium is reached from an initially non-equilibrium state, or even whether it can be reached or not, has long been a problem. A modern investigation came with the Fermi-Pasta-Ulam simulation as soon as the electric computer was available \cite{fpu}. The surprising result was that the system exhibited a long-time periodic behavior without any sign of ergodicity, which was later ascribed to the integrability of the model in the continuum limit \cite{zubusky}. More recently, the problem revived again because of the possibility of using ultracold atoms to address it experimentally \cite{weiss}. Many different models, integrable \cite{weiss,luttinger,rigol_integrable} or non-integrable \cite{manmana,kollath07,rigol_nature,rigol_09}, are investigated. For those models integrable, as expected, no thermalization, or at least no thermalization in the usual Gibbs ensemble sense is observed \cite{weiss,luttinger,rigol_integrable}. What is unexpected is that, even for some non-integrable models \cite{manmana,kollath07,rigol_09}, thermalization does not occur, at least at finite size. Moreover, even if thermalization does show up \cite{rigol_nature,rigol_09}, it thermalizes only in the sense that some physical observables relax to the predicted values of a microcanonical/canonical ensemble---yet the time-averaged density matrix itself shares little feature with a microcanonical/canonical ensemble (an exception is \cite{roux09}, where signature of this is observed). Therefore, the system thermalizes in a weak or pragmatic sense, since it is the few simple observables that are most ready to measure and thus of most concern. 

In this Letter, we investigate thermalization of the two-component Bose-Hubbard model. We find that this model, known as non-integrable, in some regime, does thermalize very well at a finite size. Remarkably, unlike previous works, it is not only some simple observables that thermalize, but also the time-averaged density matrix (of the whole system) itself, which can be well approximated by a canonical ensemble density matrix. The motivation is to simulate the everyday experience that two objects initially at different temperatures, when brought in contact, equilibrate eventually. Here the two species of bosons act as the two objects. It is assumed that initially each component is at equilibrium in themselves and at some finite but different temperatures and there is no interaction between them. Then at time $t=0$, the inter-component interaction is switched on. The subsequent evolution is studied. 

The Hamiltonian is ($\hbar=k_B=1$) $H_t=H_a+H_b+\theta_t H_{ab}$, where $H_{a,b}$ are Hamiltonians of components $a$ and $b$ respectively. Explicitly, $H_a=-J_a\sum_{m=1}^M(a^{\dagger}_ma_{m+1}+\text{h.c.})+ \frac{U_a}{2}\sum_{m=1}^M a^\dagger_m a^\dagger_m a_m a_m$ and $H_a \leftrightarrow H_b$ as $a \leftrightarrow b$. Here $M$ is the number of sites and periodic boundary condition is assumed. The inter-component interaction is of the on-site type $H_{ab}=U_{ab}\sum_{m=1}^M a_m^\dagger a_m b_m^\dagger b_m$. The control function is defined as $\theta_{t< 0}=0$ while $\theta_{t\geq 0}=1$. By assumption, the initial density matrix of the whole system is $\rho(0)=\rho_a(0)\otimes \rho_b(0)$, where the initial density matrices of the two components are ($\alpha =a,b$) $\rho_{\alpha}(0)= \frac{1}{Z_\alpha} \sum_{j=1}^{D_{\alpha}} e^{-\beta_{\alpha} E_{\alpha}^j} |j \rangle_{\alpha} \langle j| $. Here $\beta_\alpha$ is the inverse temperature of the component $\alpha$, $|j\rangle_{\alpha}$ denotes the $j$-th eigenstate of $H_{\alpha}$ with eigenvalue $E^j_{\alpha}$, and $Z_{\alpha}= \sum_{j=1}^{D_{\alpha}} e^{-\beta_\alpha E_{\alpha}^j} $ is the partition function. The dimension of the Hilbert space $\mathcal{H}_{\alpha}$ of component $\alpha$ is $D_\alpha=\frac{(M+N_\alpha-1)!}{(M-1)!N_\alpha!}$, with $N_\alpha$ being the total atom number of component $\alpha$.

Now turn on the interaction. Denote the $n$-th eigenstate (with eigenvalue $E_n$) of the final Hamiltonian $H_{t\geq 0}$ as $|\psi_n\rangle$. The density matrix at an arbitrary time later is formally $\rho(t)= \sum_{n,l=1}^D \langle \psi_n | \rho(0) | \psi_l\rangle e^{-i(E_n-E_l) t} |\psi_n \rangle \langle \psi_l |$, where $D=D_a D_b$ is the dimension of the full Hilbert space $\mathcal{H}=\mathcal{H}_a \otimes \mathcal{H}_b $. At this point the time-averaged density matrix is defined as
\begin{equation}
\bar{\rho}\equiv \lim_{\tau \rightarrow \infty} \frac{1}{\tau} \int_0 ^{\tau} d t \rho(t)=\sum_{\substack{n,l=1; E_n=E_l}}^D \langle \psi_n| \rho(0) | \psi_l\rangle  |\psi_n \rangle \langle \psi_l |. \nonumber
\end{equation}
The operator $\bar{\rho}$ is of significant relevance for our purpose for multiple reasons. First, it is observable-free. Second, the time-averaged value of an arbitrary operator is given simply by $\overline {\langle O\rangle} \equiv \lim_{\tau \rightarrow \infty} \frac{1}{\tau} \int_0 ^{\tau}  tr(\rho(t)O)d t=tr(\bar{\rho}O)$. Third, the process of averaging over time is a process of relaxation in the sense that the entropy associated with $\bar{\rho}$ is definitely no less than that with the density matrix at an arbitrary time, i.e., $S(\bar{\rho})\geq S(\rho(t))=S(\rho(0))$. This is a corollary of the Klein inequality \cite{chuang} and is reasonable since $\rho(0)$ contains all the information of $\bar{\rho}$ while the inverse is invalid. The equality also means that $\rho(t)$ will never be damped, and time-averaging is essential.

We note that $H_t$ is invariant under simultaneous translations $(a_m,b_m)\rightarrow (a_{m+1},b_{m+1})$. Especially, $H_{t<0}=H_a+H_b$ is invariant under the two translations individually. This implies the conservation of quasi-momentum(a) (QM). The QM of component $a$ is defined as $q_a=\sum_{k=0}^{M-1} k a_k^\dagger a_k \pmod{M}$, where $ a_k^\dagger = \frac{1}{\sqrt{M}} \sum_m e^{i m 2\pi k/M} a_m^\dagger $ is the creation operator of an atom in the $k$-th Bloch state. Similar operators are defined for component $b$. Our strategy is then to transform to the QM space.
We decompose the total Hilbert space $\mathcal{H}$ into $M$ subspaces according to the total QM $q=q_a+q_b$, $\mathcal{H}=\oplus_{q=0}^{M-1}\mathcal{H}^{(q)}$, which are further decomposed according to the QM of the two components $(q_a,q_b)$, $\mathcal{H}^{(q)}=\oplus_{q_a=0}^{M-1} \mathcal{H}_a^{(q_a)}\otimes \mathcal{H}_b^{(q-q_a)}$. The Hamiltonian and density matrix are always block-diagonal with respect to the $q$-subspaces, $H_t=\oplus_{q=0}^{M-1} H^{(q)}_t$ and $\rho(t)=\oplus_{q=0}^{M-1} \rho^{(q)}(t)$ \cite{ini}. In each $q$-subspace, generally there is no degeneracy between the eigenstates $\{ |\psi_n\rangle \}$, therefore, the $\bar{\rho}$ in each $q$-subspace is simply the diagonal part of the initial density matrix in the $\{ |\psi_n\rangle \}$ representation, i.e.,
$ \langle \psi_n| \bar{\rho}^{(q)}| \psi_l \rangle = \delta_{nl}\langle \psi_n| \rho^{(q)}(0)| \psi_l \rangle $. Here it is necessary to mention that for some quantities (e.g. $\langle a_k^\dagger a_k \rangle $) studied below, we should have averaged over all the $q$-subspaces. However in this paper we do not bother doing so, because the system behaves quantitatively similar in all the $q$-subspaces \cite{beta}. A single $q$-subspace captures the overall behavior well. Therefore, we shall focus on some specific $q$-subspace and take the normalization $tr(\rho^{(q)}(t))=1$.

\begin{figure}[tbh]
\centering
\includegraphics[bb=33   269   590   602, width=0.35\textwidth]{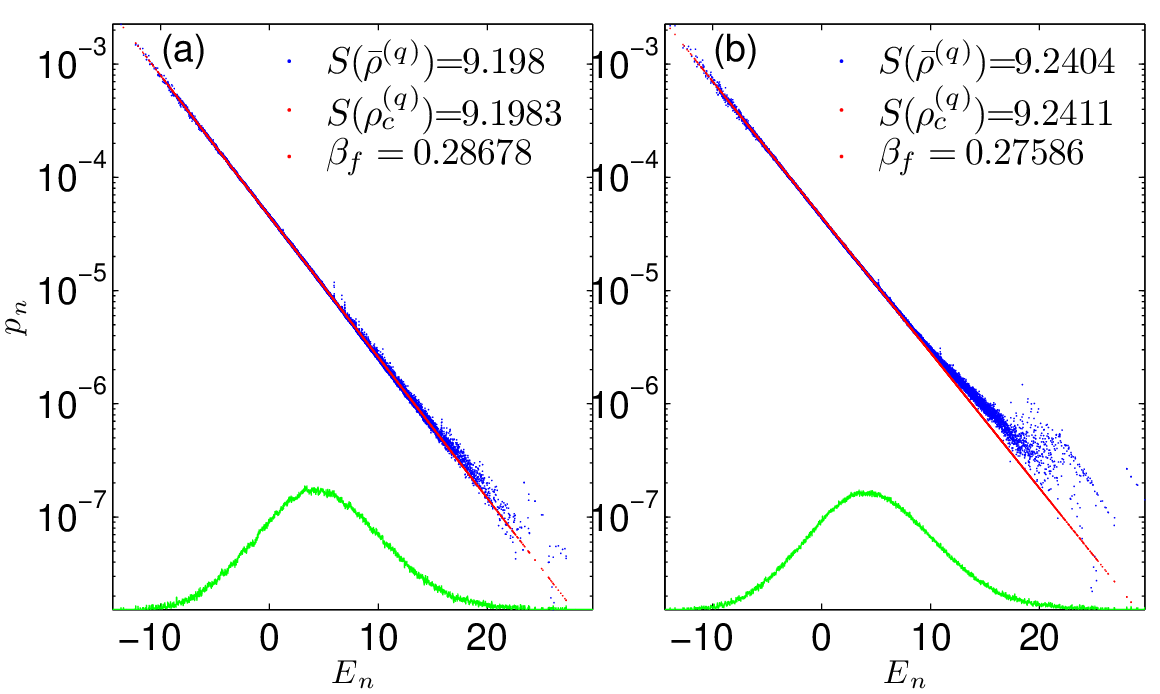}
\caption{\label{fig1} Diagonal elements of $\bar{\rho}^{(q)}$ (blue dots) and $\rho_c^{(q)}$ (red dots) versus the eigenvalues $E_n$. The parameters are $(N_a,N_b,M,q)=(4,4,9,1)$, $(J_a,J_b,U_a,U_b)=(1,1,2,2)$, $(\beta_a,\beta_b)=(0.2,0.4)$, and (a) $U_{ab}=0.5$; (b) $U_{ab}=1$. The dimension of the $q$-subspace is $D^{(q)}=27,225$. The green lines depict the coarse-grained density of states of $H^{(q)}_{t>0}$, just for reference (not corresponding to the vertical axes). The inverse temperature $\beta_f$ and the entropies of $\bar{\rho}^{(q)}$ and $\rho_c^{(q)}$ are shown in the inserts.}
\end{figure}

As mentioned above, we are motivated to study the relaxation dynamics of the initial non-equilibrium system. A natural question is then how $\bar{\rho}^{(q)}$ is like. As revealed by Fig.~\ref{fig1}, at least in some regime, it has strong characteristic of a canonical ensemble. In each panel, the occupations on the eigenstates $p_n=\langle \psi_n | \bar{\rho}^{(q)} | \psi_n \rangle$ are plotted versus the eigenvalues $E_n$. It is amazing that most of the points cling close to a straight line except at the ends of the spectrum, and the straight line is actually the prediction of a canonical ensemble $\rho_{c}^{(q)}(\beta_f)\equiv e^{-\beta_f H^{(q)}_{t>0}}/tr(e^{-\beta_f H^{(q)}_{t>0}})$ with the same energy as $\bar{\rho}^{(q)}$, i.e., $\bar{E}\equiv tr(H^{(q)}_{t>0} \bar{\rho}^{(q)})=tr(H^{(q)}_{t>0} \rho_{c}^{(q)}(\beta_f)) $. The situation improves even further if the initial temperatures $1/\beta_{a,b}$ are increased \cite{beta_infty}. Besides Fig.~\ref{fig1}, we have explored the parameter space extensively. The rule of thumb is that when the $J$'s and $U$'s are comparable, i.e., when the model is far from the integrable limits, results similar to Fig.~\ref{fig1} can occur. Of course, we should mention that the fitting is not always so good as in Fig.~\ref{fig1}. In the low temperature or unbalanced case ($|\beta_a-\beta_b|$ large), the fitting worsens. We will come back to this later.

\begin{figure}[tbh]
\centering
\includegraphics[bb=36   263   583   607, width=0.30\textwidth]{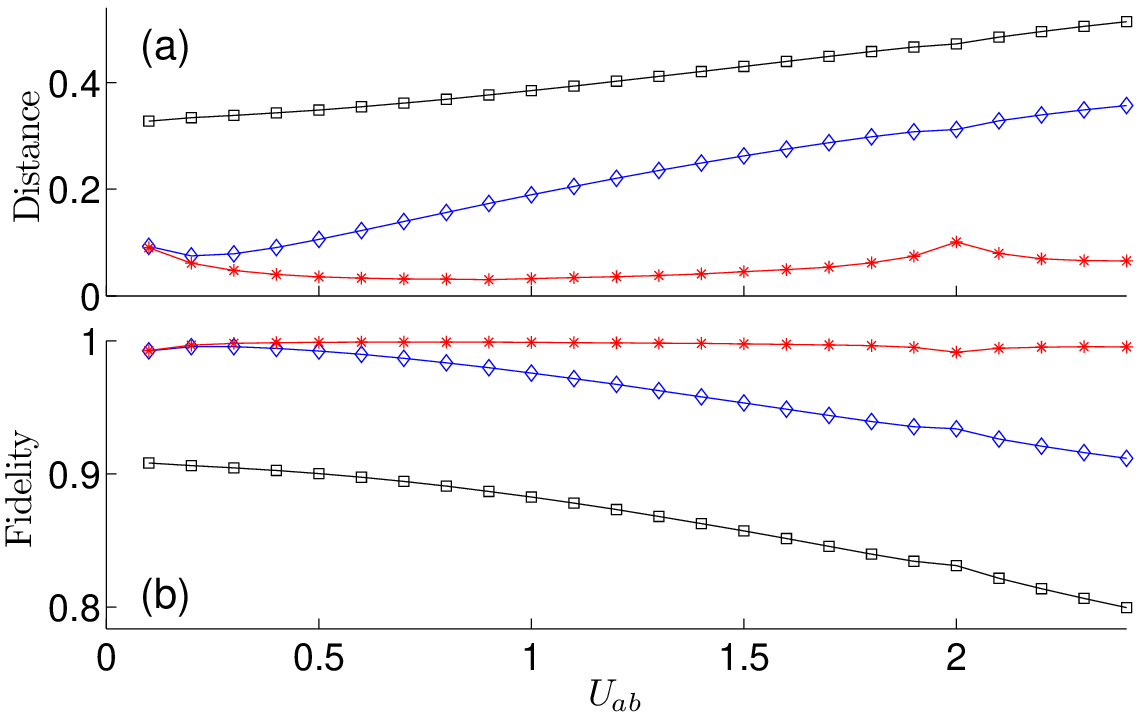}
\caption{\label{fig2} (a) Distance and (b) Fidelity between $\bar{\rho}^{(q)}$ and the canonical ensemble density matrix $\rho_c^{(q)}$ (red $*$), product density matrix $\rho_{prod}^{(q)}$ (blue $\diamond$), and the initial density matrix $\rho^{(q)}(0)$ (black $\square$). The parameters are $(N_a,N_b,M,q)=(4,4,9,1)$, $(J_a,J_b,U_a,U_b)=(1,1,2,2)$, and $(\beta_a,\beta_b)=(0.3,0.8)$.}
\end{figure}

Figure \ref{fig1} gives us an overall impression of $\bar{\rho}^{(q)}$. To characterize it further, we use the tools of \textsl{distance} and \textsl{fidelity} to study its relation to some reference density matrices. The three reference density matrices chosen are respectively the canonical ensemble one mentioned above, the product state ($q$-section actually) $\rho_{prod}^{(q)}(\beta_f)=\mathcal{N}_1\oplus_{q_a=0}^{M-1} e^{-\beta_f H_a^{(q_a)}} \otimes e^{-\beta_f H_b^{(q-q_a)}}$, and the initial density matrix $\rho^{(q)}(0)=\mathcal{N}_2\oplus_{q_a=0}^{M-1} e^{-\beta_a H_a^{(q_a)}} \otimes e^{-\beta_b H_b^{(q-q_a)}}$, where $\mathcal{N}_{1,2}$ are normalization coefficients such that $tr(\rho_{prod}^{(q)})=tr(\rho^{(q)}(0))=1$. The distance and fidelity between two density matrices are defined as $D(\rho,\sigma)=\frac{1}{2}tr\sqrt{(\rho-\sigma)^2}$ and $F(\rho,\sigma)=tr\sqrt{\rho^{1/2}\sigma\rho^{1/2}}$, respectively. They both take values in the range of $[0,1]$ and are closely related to each other by the inequality $1-F\leq D \leq \sqrt{1-F^2}$ \cite{chuang}. Two density matrices are close to each other if $D$ and $1-F$ are much smaller than unity. In Fig.~\ref{fig2}, the $D$'s and $F$'s are shown as the interaction strength $U_{ab}$ is varied while all other parameters fixed. We see that in the full range of $U_{ab}$ investigated, $\rho_c^{(q)}$ is the one closest to $\bar{\rho}^{(q)}$, while $\rho^{(q)}(0)$ is the farthest [and so is $\rho^{(q)}(t) $ actually, because $D(\rho^{(q)}(0),\bar{\rho}^{(q)})=D(\rho^{(q)}(t),\bar{\rho}^{(q)})$ and $F(\rho^{(q)}(0),\bar{\rho}^{(q)})= F(\rho^{(q)}(t),\bar{\rho}^{(q)})$ by the unitary invariance of $D$ and $F$], with $\rho_{prod}^{(q)}$ being the intermediate one. Moreover, the distance and infidelity between $\bar{\rho}^{(q)}$ and $\rho_c^{(q)}$ are much smaller than unity throughout the range. This indicates that the ``artificial'' time-averaging is very effective in thermalizing an initially non-equilibrium state. The fact that $\rho_c^{(q)}$ is always a better approximation of $\bar{\rho}^{(q)}$ than $\rho_{prod}^{(q)}$ indicates that the two subsystems equilibrate as a whole instead of factorizably. This is consistent with the fact that the inter-component interaction is a bulk type not a surface type.

\begin{figure}[tbh]
\centering
\includegraphics[bb=32   250   580   607, width=0.30\textwidth]{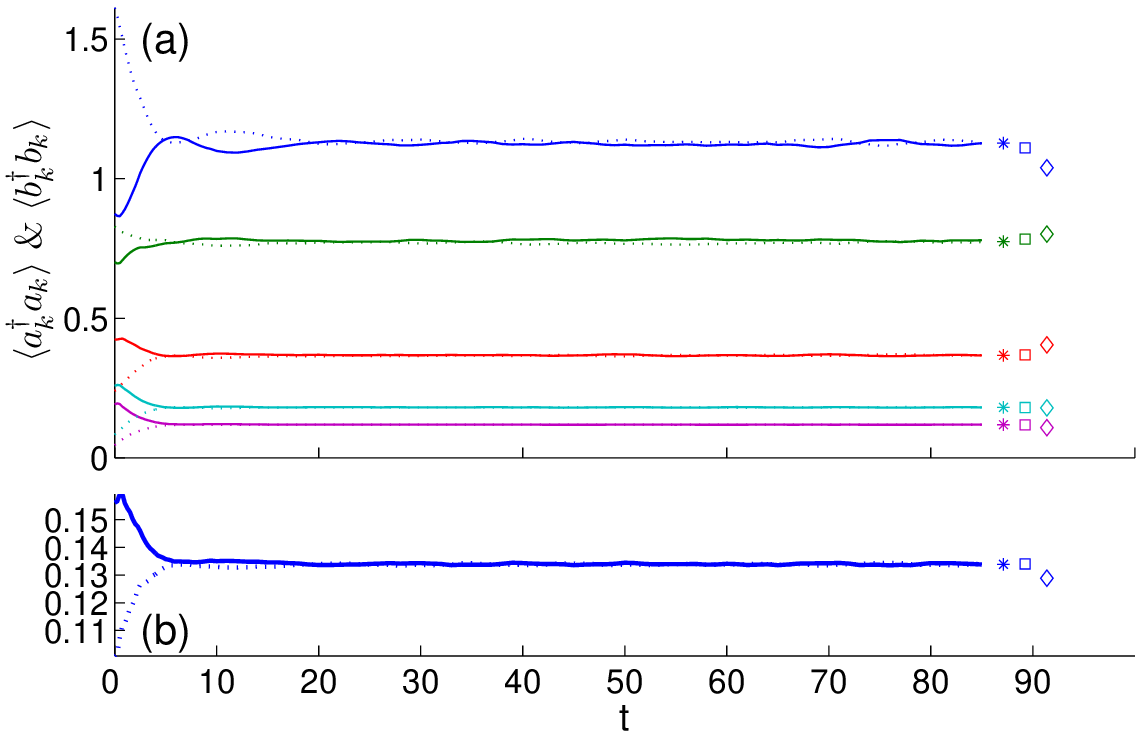}
\caption{\label{fig3} (a) Evolution of the occupations on the Bloch states, $\langle a_k^\dagger a_k \rangle$ (solid lines) and $\langle b_k^\dagger b_k \rangle$ (dotted lines). From up to down, $k=0,\ldots,4$. Other $k$'s are not shown because lines with $k$ and $M-k$ are very close to each other all the time. (b) Evolution of the on-site atom-atom correlations, $\langle a_m^\dagger a_m^\dagger a_m a_m \rangle $ (solid line) and $\langle b_m^\dagger b_m^\dagger b_m b_m \rangle $ (dotted line). For each pair of lines, the markers of the same color on the right hand side indicate the average value or $\bar{\rho}$ prediction ($*$), $\rho_c^{(q)}$ prediction ($\square$), and $\rho_{mic}^{(q)}$ prediction ($\diamond$), respectively. The parameters are the same as in Fig.~\ref{fig2} with $U_{ab}=1$.}
\end{figure}

It has been verified in many aspects that $\bar{\rho}^{(q)}$ can be well approximated by some canonical ensemble density matrix $\rho_c^{(q)}(\beta_f)$. We then anticipate that the time-averaged values (predicted by $\bar{\rho}^{(q)}$) of many physical quantities of interests are also well predicted by $\rho_c^{(q)}$. Moreover, if dephasing is effective, we shall see perfect relaxation phenomenon in the dynamics of the physical quantities. It is indeed the case. As shown in Fig.~\ref{fig3}, the occupations on the Bloch modes $\langle a_k^\dagger a_k \rangle$ [and $\langle b_k^\dagger b_k \rangle$] relax to their average values quickly, exhibiting minimal fluctuations \cite{peter}, and these values are very close to those predicted by $\rho_c^{(q)}$ (but with much larger errors to those by the microcanonical ensemble density matrix $\rho_{mic}^{(q)}$ \cite{mic}). That is, these quantities relax and relax to their equilibrium values in $\rho_c^{(q)}$. Note that due to the symmetric parameters chosen, $\langle a_k^\dagger a_k \rangle=\langle b_k^\dagger b_k \rangle$ for $\bar{\rho}^{(q)}$, $\rho_c^{(q)}$, and $\rho_{mic}^{(q)}$ all, and indeed we see in Fig.~\ref{fig3} that $\langle a_k^\dagger a_k \rangle$ and $\langle b_k^\dagger b_k \rangle$ merge from distinct initial values. Besides these single-particle quantities, also shown are the evolution of the two-particle quantities $\langle a_m^\dagger a_m^\dagger a_m a_m \rangle $, and $a \leftrightarrow b$, with similar behavior observed. 
\begin{figure}[tbh]
\centering
\includegraphics[bb=20 302 596 590, width=0.30\textwidth]{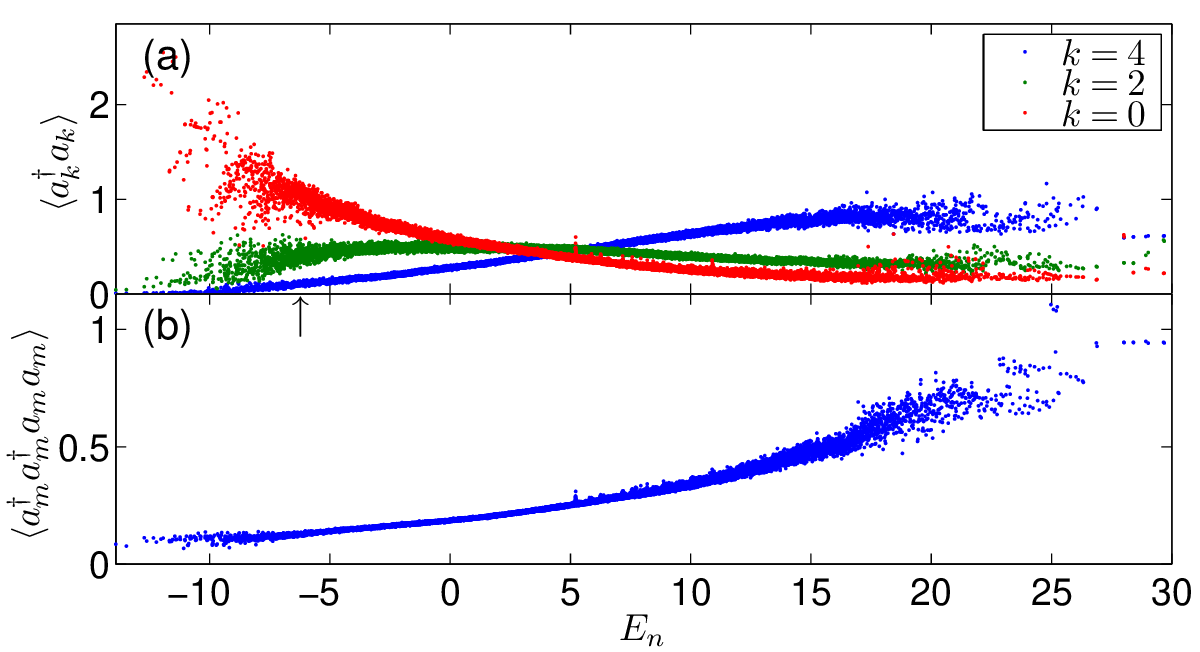}
\caption{\label{fig4}  (a) Occupations on the Bloch states (one-particle operators) $\langle a_k^\dagger a_k \rangle$ and (b) on-site atom-atom correlation (two-particle operator) $\langle a_m^\dagger a_m^\dagger a_m a_m \rangle $ for each eigenstate of ${H}^{(q)}_{t>0}$. The $\uparrow$ indicates the average energy $\bar{E}$ in Fig.~\ref{fig3}. The parameters are the same as in Fig.~\ref{fig3}.}
\end{figure}

Here some remarks about the connection between thermalization of the density matrix and that of physical observables are in order. The point is that the former implies, but is not necessary for, the later. Two density matrices can yield the same expectation values for a few ``realistic'' physical quantities, yet be quite far apart in terms of $D$ and $F$. Actually, it is common knowledge that in the thermodynamic limit, for a generic system, the predications of a micro-canonical ensemble and a canonical one agree well, yet it is easy to persuade oneself that the distance and fidelity between the corresponding density matrices are $(1-D,F) \ll 1$. The reason is formulated as the eigenstate thermalization hypothesis (ETH) \cite{eth}, which is verified in some finite systems \cite{rigol_nature,rigol_09}. According to ETH, the expectation value of a generic few-body physical quantity varies little between eigenstates close in energy, therefore, the detailed distribution $p_n$ does not matter as long as it is narrow in energy. Here it is verified that ETH is acceptable for the variables in Fig.~\ref{fig3} (see Fig.~\ref{fig4}). However, it plays a marginal role in the thermalization there. As shown in Fig.~\ref{fig4}, the average energy $\bar{E}$ falls at the head of the spectra where ETH is not so good. Thus we see in Fig.~\ref{fig3} that the predictions of $\rho_{mic}^{(q)}$ deviate significantly from the true values, yet the predictions of $\rho_c^{(q)}$ agree much better with $\bar{\rho}^{(q)} $. The situation persists in a wide range of parameters even if $\bar{E}$ falls in the body of the spectra where ETH is good. It is thus apparent that it is the detailed distribution, which is more accurately captured by $\rho_c^{(q)}$ than $\rho_{mic}^{(q)}$, that really matters. However, this does not rule out the possibility that in the thermodynamic limit, the distribution of $E$ falls in a small interval where ETH holds and thus both $\rho_c^{(q)}$ and $\rho_{mic}^{(q)}$ agree with $\bar{\rho}^{(q)}$ as for the physical observables.

\begin{figure*}[tb]
\begin{minipage}[b]{0.30 \textwidth}
\centering
\includegraphics[bb=20   253   574   604, width=\textwidth]{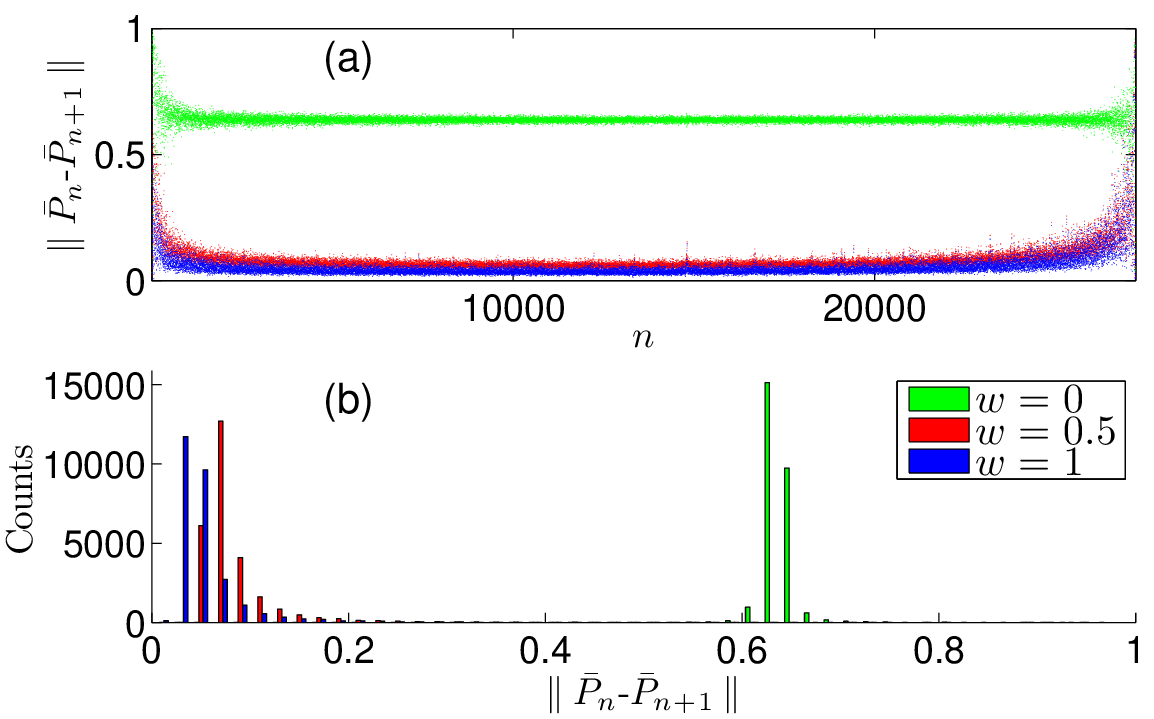}
\end{minipage}
\begin{minipage}[b]{0.30 \textwidth}
\centering
\includegraphics[bb=47   263   569   594, width=\textwidth]{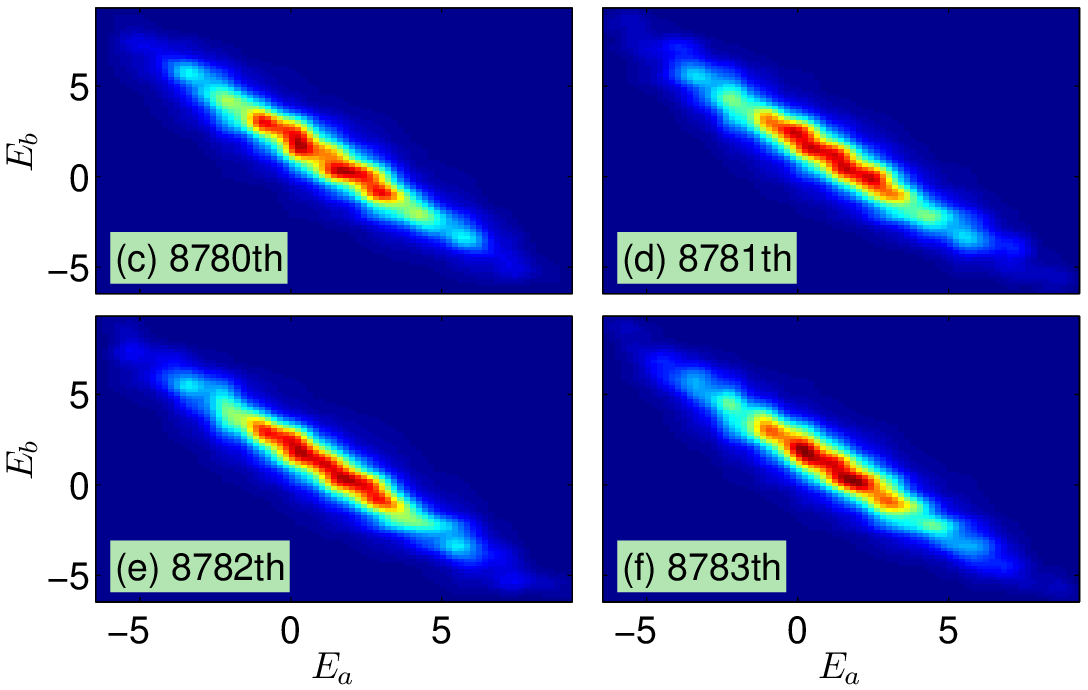}
\end{minipage}
\caption{\label{fig5} (a) Distances between the coarse-grained probability distributions corresponding to adjacent eigenstates, with Gaussion broadening. The green, red, and blue dots correspond to $w=0$, $0.5$, and $1$, respectively. (b) Histogram of $\parallel \bar{P}_n-\bar{P}_{n+1} \parallel$ in bins of length $0.02$. (c)-(f) illustrate the Gaussion broadened ($w=0.5$) probability distributions corresponding to four successive eigenstates. The distances between adjacent pairs are $(0.0460,0.0419,0.0502)$. The parameters are the same as in Fig.~\ref{fig1}b (but note that $\beta_{a,b}$ are irrelevant here).}
\end{figure*}

We now return to Fig.~\ref{fig1}. The fact that the occupations on the eigenstates $
p_n=\mathcal{N} \sum_{ |ij\rangle \in \mathcal{H}^{(q)} } e^{-\beta_a E^i_{a}-\beta_b E^j_{b}} |\langle ij | \psi_n \rangle|^2
$, where $\mathcal{N}$ is a normalization factor, are well captured by the formula $p_n\propto e^{-\beta_f E_n}$, is too remarkable to be overlooked \cite{indirect}. This fact is non-trivial since at the first glance there is no clue in the expression. So far we have not understood it fully but we do understand the weak fact that $p_n/p_m\simeq 1 $ if $|E_n- E_m| \ll 1/\beta_f$, in particular, $p_n/p_{n+1}\simeq 1$. We have
$ p_n=\mathcal{N}  \iint_{-\infty}^{+\infty} d E_a dE_b P_n(E_a,E_b) e^{-\beta_a E_a-\beta_b E_b} $,
with the probability distribution function $P_n(E_{a},E_{b}) =\sum_{ |ij\rangle  } |\langle ij | \psi_n \rangle|^2  \delta(E_{a}-E^i_{a},E_{b}-E^j_{b})$ \cite{probability}. It consists of a series of $\delta$-functions with fixed positions but $n$-dependent amplitudes and is an intrinsic property of $|\psi_n\rangle$ in terms of $|ij\rangle$. Coarse-graining $P_n$ by replacing the $\delta$-functions with some regular peaked function $f(x,y)$ (satisfying $\iint dx dy f=1$ and $f>0$), we rewrite $p_n$ as 
\begin{equation}\label{pn}
p_n=\frac{\mathcal{N}}{c}\iint_{-\infty}^{+\infty} d E_a dE_b \bar{P}_n(E_a,E_b) e^{-\beta_a E_a-\beta_b E_b}.  
\end{equation}
Here the coarse-grained distribution $\bar{P}_n(E_a,E_b) =\sum_{ |ij\rangle  } |\langle ij | \psi_n \rangle|^2 f(E_a-E^i_a,E_b-E^j_b)$, and the constant $c=\iint dx dy e^{-\beta_a x -\beta_b y} f(x,y) $, which is $n$-independent. The fact that $\bar{\rho}^{(q)}$ and $\rho_c^{(q)}$ in Fig.~\ref{fig1} always agree well in the high temperature regime $\beta_{a,b}\leq 0.4$ suggests that there exists some $f$ such that for most $n$'s, $\bar{P}_n$ and $\bar{P}_{n+1}$ are close to each other in a certain sense---an intrinsic property independent of $\beta_{a,b}$. As a measure of the difference between two probability distributions, we have the metric $ \parallel \bar{P}_n-\bar{P}_m \parallel=\frac{1}{2} \iint d E_a d E_b |\bar{P}_n-\bar{P}_m| $ \cite{chuang}. By this metric, two distributions are close to each other if $\parallel \bar{P}_n-\bar{P}_m \parallel \ll 1$. We have studied the distances between $\bar{P}_n$ and $\bar{P}_{n+1}$ using the Gaussion function $f(x,y)=\frac{1}{\pi w^2}\exp[-(x^2+y^2)/ w^2]$, where $w$ is the adjustable width. The results are shown in Fig.~\ref{fig5}. We see that although for the initial distributions ($w=0$, in this case $\bar{P}_n$ degenerates to $P_n$), $\parallel \bar{P}_n-\bar{P}_{n+1} \parallel$ centers around $0.63$, once broadening is triggered, it shrinks abruptly. For $w=0.5$ and $1$, over $84\%$ and $93\%$ pairs have a distance less than $0.1$ respectively. Moreover, those pairs having large distances fall mainly at the ends of the spectrum, consistent with the fact that in Fig.~\ref{fig1} the fitting is bad at the ends (large fluctuations). In Figs.~\ref{fig5}c-f, the broadened distributions $\bar{P}_n(E_a,E_b)$ for four successive $n$'s are illustrated. It is apparent that they agree even in details. We also observe that the contour of $\bar{P}_n$ stretches along the direction $E_a + E_b =\text{const}$. This is reasonable since $H_{ab}$, as a perturbation, mixes eigenstates of $H_a + H_b$ with adjacent eigenenergies best. At this point, we can understand why the fitting in Fig.~\ref{fig1} is good and why low temperature and large difference $| \beta_a-\beta_b |$ are two adverse conditions for the fitting. The exponential weight function $e^{-\beta_a E_a -\beta_b E_b}$ descends fastest in the direction $(\beta_a,\beta_b)$. In the two adverse conditions, the weight function changes significantly across the region where $\bar{P}_n$ takes significant values and which extends primarily in the direction $(1,-1)$. This non-uniformity potentially will spoil the closeness between $\bar{P}_n$ and $\bar{P}_{n+1}$ in terms of the metric above. 

Finally, we should mention that the essence of coarse-graining is to smear out the irrelevant details of the distribution $P_n$ retaining only the relevant overall information. As the size of the system grows, the number of $\delta$-functions within the radius $w$ increases exponentially and the coarse-graining shall be even more effective in reducing the distance between $P_n$ and $P_{n+1}$, and therefore it is legitimate to expect even better fitting.

To conclude, it is demonstrated that the generic two-component Bose-Hubbard model exhibits perfect thermalization. It is strong thermalization in that not merely the average values of a few physical variables but even the time-averaged density matrix itself thermalizes. We also note that our scenario is potentially realizable with cold atoms in optical lattices. Atoms in optical lattices are at finite temperatures necessarily \cite{zhou} and the inter-component interaction can be controlled with Feshbach resonance. Moreover, the ensemble average is offered automatically by a two-dimensional array of one-dimensional optical lattices. 

We are grateful to L. M. Duan for helpful suggestions.


\begin{thebibliography}{99}
\bibitem{ini}
In particular, the initial density matrix can be decomposed finer as $\rho^{(q)}(0)=\oplus_{q_a=0}^{M-1} \rho_a^{(q_a)}(0) \otimes \rho_b^{(q-q_a)}(0) $ with $\rho_\alpha^{(q_\alpha)} \in \mathcal{H}_\alpha^{(q_\alpha)}$, since $H_{t<0}^{(q)}$ can be decomposed similarly.
\bibitem{beta_infty}
This makes sense since in the extremal and trivial case of $\beta_{a,b}=0$, $\rho_{a,b}(0)=I_{a,b}$ and therefore $\rho(0)=I=\bar{\rho}$, which is fitted exactly by a canonical ensemble with $\beta_f=0$.
\bibitem{indirect}
This fact is consistent with the other fact that $S(\bar{\rho}^{(q)})$ is very close to $S(\rho_c^{(q)})$, as shown in the inserts of Fig.~\ref{fig1}. Since the canonical distribution is the one and the only one that maximizes the entropy when the average energy is fixed [see E.~T. Jaynes, Phys. Rev. \textbf{106}, 620 (1957)], $\bar{\rho}^{(q)}$ should be close to $\rho_c^{(q)}$ if $S(\bar{\rho}^{(q)}) \lesssim S(\rho_c^{(q)})$.
\bibitem{probability}
It is so called because $\iint d E_a d E_b P_n(E_a,E_b)=1$ and $P_n(E_a,E_b)\geq 0$.
\end{thebibliography}
\end{document}